\pdfoutput=1
\documentclass[11pt]{article}

\usepackage[preprint]{acl}
\usepackage{booktabs}
\usepackage{times}
\usepackage{latexsym}

\usepackage[T1]{fontenc}
\usepackage[utf8]{inputenc}

\usepackage{microtype}

\usepackage{inconsolata}

\usepackage{graphicx}

\title{Human-Centered Design Recommendations for LLM-as-a-Judge}

\author{Qian Pan \and Zahra Ashktorab \and Michael Desmond \and Martin Santillan Cooper \\ {\bf James Johnson \and Rahul Nair \and Elizabeth Daly \and Werner Geyer}  \\
       IBM Research}
\author{Qian Pan \\ \texttt{Qian.Pan@ibm.com}\\ IBM Research \\  Cambridge, MA, USA 
  \And  Zahra Ashktorab \\
  \texttt{zahra.ashktorab1@ibm.com} \\ IBM Research\\ Yorktown Heights, NY, USA
  \And  Michael Desmond \\ \texttt{mdesmond@us.ibm.com} \\ IBM Research\\ Yorktown Heights, NY, USA 
        \AND
        Martin Santillan Cooper \\ \texttt{msantillancooper@ibm.com}\\ IBM Research \\  Capital Federal, Argentina 
  \And  James Johnson \\
  \texttt{jmjohnson@us.ibm.com} \\ IBM Research\\ Cambridge, MA, USA 
  \And  Rahul Nair \\ \texttt{rahul.nair@ie.ibm.com} \\ IBM Research\\ Mulhuddart, Dublin, Ireland 
         \AND
        Elizabeth Daly \\ \texttt{elizabeth.daly@ie.ibm.com}\\ IBM Research \\  Mulhuddart, Dublin, Ireland 
  \And  Werner Geyer \\
  \texttt{werner.geyer@us.ibm.com} \\ IBM Research\\ Cambridge, MA, USA 
 }

\begin{document}
\maketitle
\begin{abstract}
%Advancements in large language models (LLMs) complicate text generation quality evaluation. 
Traditional reference-based metrics, such as BLEU and ROUGE, are less effective for assessing outputs from Large Language Models (LLMs) that produce highly creative or superior-quality text, or in situations where reference outputs are unavailable. While human evaluation remains an option, it is costly and difficult to scale. Recent work using LLMs as evaluators (LLM-as-a-judge) is promising, but trust and reliability remain a significant concern. Integrating human input is crucial to ensure criteria used to evaluate are aligned with the human's intent, and evaluations are robust and consistent. This paper presents a user study of a design exploration called EvaluLLM,  that  enables users to leverage LLMs as customizable judges, promoting human involvement to balance trust and cost-saving potential with caution. Through interviews with eight domain experts, we identified the need for assistance in developing effective evaluation criteria aligning the LLM-as-a-judge with practitioners' preferences and expectations. We offer findings and design recommendations to optimize human-assisted LLM-as-judge systems.
 % Rapid advancements in large language model (LLM) technology have complicated the evaluation of text generation quality. Traditional metrics like BLEU and ROUGE are less effective for assessing outputs from LLMs that produce highly creative or superior-quality text. While human evaluation remains reliable, it is costly and difficult to scale. Recent efforts focus on using LLMs as customizable judges for natural language generation (NLG), showing promising results. This paper describes EvaluLLM, a tool that allows practitioners to use an LLM as a customizable NLG judge. Through interviews with eight domain experts using the tool, we found that practitioners need help developing effective evaluation criteria that align the LLM judge with their preferences and expectations. We offer findings and design recommendations for optimizing LLM-as-judge systems.
\end{abstract}

\section{Introduction} 
%\textcolor{red} {Will circle back to change introduction, add citations,etc}
Recent advancements in Large Language Models (LLMs) challenge traditional methods of assessing natural language generation (NLG) quality, as known metrics, such as BLEU \cite{papineni2002bleu} and ROUGE \cite{lin2004rouge}, fall short for creative tasks. The diverse and expanding capabilities of LLMs \cite{liang2022holistic} present a selection challenge for practitioners, requiring evaluations of extensive outputs across contexts like summarization and retrieval-augmented generation (RAG). The subjective and use case-specific nature of emerging NLG tasks often demands human review, making the evaluation process hard to scale without suitable automatic metrics. While experts can perform evaluations, this is costly and impractical for rapid iteration in early development stages. \cite{gehrmann2023repairing}.

One potential solution to these challenges is to leverage the capabilities of LLMs to aid in the evaluation process. Despite not always being accurate, LLMs have the potential to significantly reduce the workload by identifying outputs where they are not confident, thus indicating where human  input may be required. Additionally, LLMs can assist practitioners in identifying and customizing criteria specific to their use case—such as, for example, faithfulness to contextual information,  naturalness of the conversation, and succinctness—with which they wish to conduct their evaluations. This customization enables a more targeted and effective assessment of model outputs, tailored to the specific requirements of their tasks. In this paper, we present results from a user study of EvaluLLM \cite{10.1145/3640544.3645216}, a tool designed to facilitate the evaluation of model outputs. EvaluLLM simplifies how practitioners choose LLMs by offering a quick way to assess and compare their performance across various tasks. This method accelerates the development of evaluation criteria and helps manage the growing variety and capabilities of LLMs. 

To understand the challenges and user needs in model evaluation that leverage LLM-as-a-Judge to automate the process, we conducted formative, semi-structured interviews with 8 practitioners (data scientists, software engineers, and AI engineers) who have been involved in model performance evaluation projects over the past year. Our interviews revealed various challenges and needs. For instance, practitioners highlighted the necessity for rapid performance comparison across different setups, the importance of defining evaluation criteria (e.g., structured and customizable templates aligned with specific use cases), and strategies for effectively integrating LLM-as-a-Judge into their workflow (e.g., starting with a small subset of data before scaling up).
In this paper, we present the following contributions:

% EvaluLLM streamlines the selection process for practitioners by providing an efficient means to assess and compare the performance of various LLMs across different generative tasks. This approach not only aids in the rapid iteration of evaluation criteria during the early development stages but also helps in managing the challenges posed by the expanding diversity and capabilities of available LLMs.
% Human evaluation is expensive and challenging to scale. Recent efforts have concentrated on leveraging LLMs as customizable NLG judges, yielding promising initial results.

\begin{itemize}
\item We describe EvaluLLM \cite{10.1145/3640544.3645216}, 
an LLM-Assisted evaluation tool that enables users to select multiple models, define custom metrics for NLG evaluation, and review the results while providing feedback to observe the agreement between human and AI evaluations.
\item We present qualitative findings from interviews with domain experts (N = 8) revealing challenges and user needs for model evaluation workflows including LLM-as-a-judge.
\item We make design recommendations and provide example feature designs to enable users to define criteria interactively, ensuring transparent and rapid access to LLM-as-a-judge's preferences while balancing trade-offs across multiple dimensions in a self-consistent manner.
\end{itemize}
%
%\begin{description}
%item \textbf{RQ1}: What are the challenges and user needs in model evaluations, and how can an evaluation %framework be designed to address these and meet user requirements?
%\item \textbf{RQ2}: What is the optimal workflow for human and LLM-as-a-Judge collaboration to effectively align %preferences?
%\end{description}

% We report on several themes that emerged through these user studies. These include the kinds of challenges and needs users might have for the LLM judge use case (e.g.,Support Rapid Performance Comparison with different configs), needs for criteria definition (e.g., Desire structured and customizable templates that align with specific use cases),
% how to effectively integrate LLM judge as part of the workflow (e.g., Begin with small subset of data, then expand ).

% In the discussion, we reflect on the outlined limitations of our design recommendations and outline future work and design directions to further validate our design recommendations and improve writing.

\section{Related work}
% \textcolor{red}{ADD MORE HERE, lacking in experience site/frond end related works}
%This work supports designing effective LLM judges for new generative tasks by using human feedback to calibrate quality and trustworthiness. We review literature on LLMs as judges and interactive evaluation to inform this area.
%This work aims to support the design of effective LLM judges for novel generative tasks by incorporating human-in-the-loop feedback for quality and trustworthiness calibration. To understand this domain, we review literature in three key areas: LLMs as judges, evaluation prompt design, and interactive evaluation.

%\subsection{LLMs as Judges} 
LLMs trained to follow instructions can generate results that surpass the quality of data produced by humans. This makes it increasingly challenging to assess the quality of natural language generation (NLG) outputs \cite{liang2022holistic} \cite{xiao2023evaluating} \cite{liu2023calibrating}. Traditional reference-based metrics, such as ROUGE \cite{lin2004rouge} and BLEU \cite{papineni2002bleu}, might not effectively capture the essence of LLM outputs, especially in scenarios where the output space is broad and varied. This means multiple different outcomes can all be valid, making it nearly impossible to create sufficiently comprehensive reference sets. Consequently, these metrics may not be reliable indicators of NLG output quality, as they often demonstrate a low correlation with human judgments \cite{freitag2022results}.

Recent advances highlight LLMs' potential as customizable judges, \cite{liu2023gpteval} \cite{wang2023chatgpt} \cite{zheng2023judging} capable of adapting to various tasks beyond traditional evaluation methods. Techniques like G-Eval \cite{liu2023gpteval} use chain-of-thought prompting and form-filling to assess NLG quality, while GPTScore \cite{fu2023gptscore} evaluates using conditional token probabilities, enhancing scoring granularity. AlpacaEval \cite{li2023alpacaeval} \cite{yuan2024self} compares model win rates, and Prometheus \cite{kim2023prometheus} is a fine-tuned LLM specifically designed for evaluation tasks. These methods align closely with human preferences, especially in creative tasks, emphasizing LLMs' ability to mimic human judgment. Their effectiveness relies on tailored prompt design and user-defined criteria for precise evaluations. 
While not part of this paper, in our own work, we have also done comprehensive benchmarking of human agreement of different LLM-as-a-judge approaches for different use cases and we found that depending on use case, LLMs as judges, and judging approach, we were able to achieve good results. Note that this is often a hard problem for humans too and inter-rater reliability can be a good reference.

Previous research has investigated using expert-labeled data to develop custom evaluation metrics like AUTOCALIBRATE \cite{liu2023calibrating}, but this method is limited by the availability of such data. For reference-free evaluations, interactive human involvement is preferable, allowing users to refine criteria effectively by reviewing outputs. ConstitutionMaker \cite{Petridis2023ConstitutionMaker} enables feedback on model outputs to iteratively refine prompts, focusing more on AI prototyping than evaluation. Other tools like Zeno \cite{cabrera2023zeno}, the What-If Tool \cite{wexler2019if}, and Errudite \cite{wu2019errudite} help identify model vulnerabilities by analyzing specific data segments. 
% Additionally, tools such as semi-auto labeling \cite{desmond2021semi} and auto-conflict resolution \cite{brachman2022reliance} support users in refining auto-labeling through active feedback. 
EvalLM \cite{Kim2023EvalLM} allows users to define criteria interactively, using LLM-as-a-judges for output ratings, although this can be limited by LLM reasoning capabilities \cite{zheng2023judging}. Our study builds on these insights, proposing a system where practitioners define criteria in natural language for LLMs to perform pairwise comparisons, enhancing trust through a "human-in-the-loop" blind review process that eliminates the need for expert data.
\section{EvaluLLM}
To explore how to support users in developing their own custom evaluation criteria for accurate and reliable evaluations that align with human preferences in a trustworthy manner, we designed and deployed EvaluLLM \cite{10.1145/3640544.3645216}. This tool enables users to generate evaluation outputs by providing a prompt, selecting multiple models, and defining LLM-as-a-Judge with custom metrics using natural language. Users can then review the results and provide feedback, inspecting the agreement between human and AI evaluations through a blind review process. In this paper, we use EvaluLLM as a conceptual design probe with users to explore the design space of how to support development of custom evaluation criteria for accurate and reliable evaluations that align with human preferences in a trustworthy manner.

% In this paper, we build upon prior work by \cite{10.1145/3640544.3645216}, in which they introduce a tool that enables users to generate evaluation outputs by providing a prompt, selecting multiple models, and defining LLM-as-a-Judge with custom metrics using natural language. In our paper, we will refer to this tool as EvaluLLM. We evaluate EvaluLLM with users to explore how to support development of custom evaluation criteria for accurate and reliable evaluations that align with human preferences in a trustworthy manner.

\begin{figure*}[t]
\centering
\includegraphics[width=1.0 \textwidth]{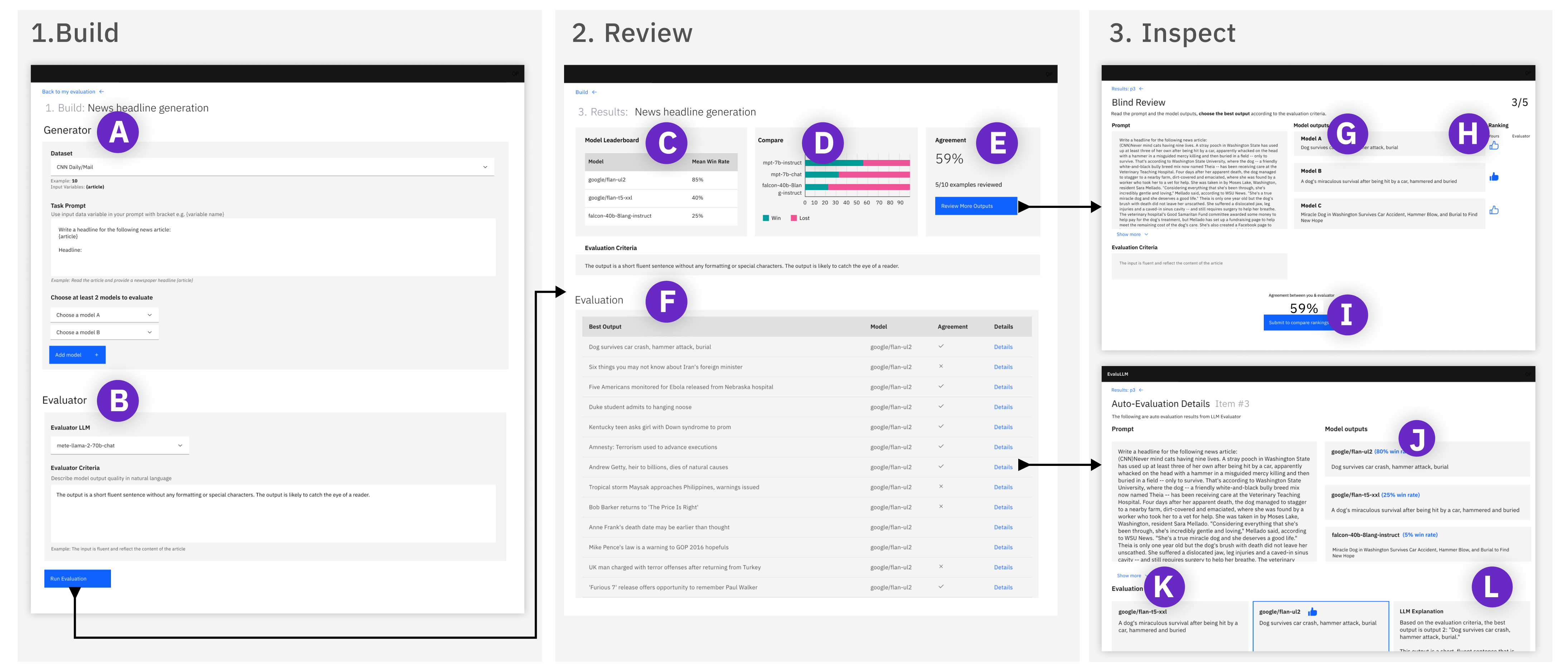}
\caption{EvaluLLM interfaces and key features}
\label{fig1}
\end{figure*}

%This Users can then review the results and provide feedback, inspecting the alignment between human and AI evaluations through a blind review process. 
%\subsection{Evaluation Workflow}
The overall user flow of EvaluLLM comprises of three stages (see Figure \ref{fig1}). The build experience focuses on defining the LLM-assisted evaluation experience to initiate the auto-evaluation process, the review experience, providing a high-level summary of the evaluation results, and the inspect experience allows users to manually examine the generated outputs through a blind review process. The data generated from this process can be used to calculate the agreement rate, assisting practitioners in better assessing the agreement between human and LLM-as-a-judges. This assessment is crucial for calibrating trust and aids in making informed decisions about whether to change configurations and rerun the evaluation.

% The EvaluLLM interface offers two user experiences: the build experience, focused on setting up and monitoring LLM-assisted evaluation, and the review experience, for exploring and assessing results 

% In the build experience, we allow users customize the generator (task prompt editing and LLM selection) and the judge (custom evaluation criteria with freeform input box). After evaluation, users move to the review experience, featuring a model leaderboard sorted by win rate. Users can manually review and rate some outputs to compare with LLM-based evaluations, producing an agreement score. This score indicates alignment between user and judge LLM on output quality, with the intent to help user to assess the correct interpretation of evaluation criteria. Users can also browse a list of evaluated instances and preference rationales provided by LLM judge. 

%\subsection{Evaluation Algorithm Considerations}
% EvaluLLMis an web based tool for assessing and comparing the effectiveness of various Language Learning Models (LLMs) on a user-defined Natural Language Generation (NLG) task. This task is specified by an input dataset $D$ and a task prompt $P$, where variables from $D$'s schema can be integrated into $P$ using curly brackets, along with fixed instructions for the task. During execution, these variable references in the prompt are substituted with actual data values from instances in $D$. The NLG task is then assessed across multiple LLMs $M$, utilizing an evaluating LLM $E$ and a set of evaluation criteria $C$ defined by the user in natural language.

In the absence of reference data, related studies suggest that LLMs may not be entirely suitable for use as numerical judges \cite{zheng2023judging}. This is because grading based on single answers may fail to detect minor distinctions between specific pairs. Furthermore, the outcomes could become unreliable, as absolute scores tend to vary more than relative pairwise results when there are changes in the judging model \cite{zheng2023judging}. To mitigate these challenges, EvaluLLM uses a pairwise comparison approach, as it can reduce the complexity of the evaluation task by breaking down the comparison of multiple outputs into smaller decisions between pairs of data which might yield to more accurate evaluation results at the cost of additional inference operations. The evaluation method involves making pairwise comparisons between the outputs of different models, similar to the AlpacaEval approach \cite{li2023alpacaeval}. However, instead of comparing outputs to a single reference, they are compared against one another. 

% Each instance $d_{i}\in D$ from the dataset, the task prompt $P$ is applied, and then then inferenced on each LLM $m_{i}\in M$ generating a series of outputs outputs $O_{i}$ for each input example $d_{i}$. The evaluation of these outputs involves a custom prompt that includes an \textit{instruction} ($P$ populated with instance variables from $d_{i}$), a pair of selected outputs $o_{1}, o_{2}\in O_{i}$, and the evaluation criteria $C$. The judge LLM is prompted to select the best output based on the evaluation criteria $C$, with its first non-space character indicating this preference. The judge also provides an explanation, aiding in human review and debugging.

% Each evaluation involves $\binom n2$  trials of comparison, where $n$ is equal to the number of generated outputs for a given data example $d_{i}\in D$. The outcomes are quantified by win rates, both locally (per instance) and globally (per model), reflecting the proportion of times an output or model was preferred in its evaluations. This win rate helps in ranking outputs for review and models for overall performance, offering an intuitive metric for users.

%\subsection{Interface Walk-through}
\subsection{Build} The build experience (see Figure \ref{fig1}) includes two major components: the Generator (Figure \ref{fig1}A) and the Evaluator (Figure \ref{fig1}B).
The Generator section (Figure \ref{fig1}A) is designed to produce evaluation data, supporting users in selecting a pre-uploaded dataset and inputting their task prompts. Users can incorporate data variables from the dataset's structure into the task prompt using the conventional curly bracket format. Additionally, the system provides a range of LLMs for users to choose from for the purpose of performance evaluation. The Evaluator section (Figure \ref{fig1}B) is where users can choose the LLM-as-a-judge model for automatic evaluation and specify the custom metrics that the judge will use to assess the outputs from the generator. This initial version of EvaluLLM, deliberately provides only a freeform input box to support maximum creativity, as the aim was to gain more insights into the types of inputs users would provide to define criteria in natural language and the kind of support users might need to define custom metrics. Once the user completes the setup, they can click the "Run Evaluation" button to initiate the evaluation.
% process using selected LLM-as-a-judge.

\subsection{Review}
Upon completion of the automatic evaluation, results are available for review. Users can view a high-level performance summary and a detailed results table. The summary includes a model leader board (Figure \ref{fig1}C), ranking selected LLMs by their win rates derived from evaluated output pairs. The performance visualization (Figure \ref{fig1}D) shows detailed win-loss statistics for each model based on pairwise comparisons by the LLM-as-a-judge. Additionally, the agreement rate (Figure \ref{fig1}E) indicates the alignment between human and LLM-as-judges, helping users gauge the reliability of evaluations. This feature becomes available after users manually rate output samples.

\subsection{Inspect}
Users can examine auto-evaluation results through two main methods. First, users can conduct a blind review, manually inspecting data to assess the reliability of LLM evaluations (Figure \ref{fig1}G). In this process, models' names are hidden to prevent bias, and users select the best output from all presented outputs. Ratings from this process are used to calculate an agreement score, which measures alignment between user and LLM-as-a-judge preferences (Figure \ref{fig1}E, I). After rating, users can view model identities and the updated agreement score (Figure \ref{fig1}H, I), providing insight into the effectiveness of the evaluation criteria. Users can also access detailed results on the review page, which displays the LLM-as-a-Judge's aggregated rankings and win rates from pairwise comparisons (Figure \ref{fig1}J). Evaluation rationales are provided next to each comparison result (Figure \ref{fig1}L, K), helping users decide whether to trust the results or adjust settings for a reevaluation.

\section{Methodology}
Our goal was to explore the challenges users encounter during LLM-assisted model evaluations and, based on our observations, to design a framework that meets their needs and supports effective collaboration between humans and LLM-as-a-judges. We used EvaluLLM to facilitate the creation of evaluation tasks and conducted our research through semi-structured interviews using Webex. Participants accessed a prototype of EvaluLLM, shared their screens, and used think-aloud methods to create evaluation tasks. Each participant worked on the same task: using an LLM-as-a-judge to identify the best model for generating headlines from the CNN/Daily Mail dataset.
%\subsection{Formative Study}
% \subsection{Overview}

%In this study, we set out to understand the challenges and users face during model evaluations and how we can design an evaluation framework that addresses the challenges and meets the needs of users during model evaluations. We also wanted to see what is the desired human and LLM-as-a-Judge collaboration workflow and how we can effectively support preference alignment between human and LLM-as-a-Judge. With these goals in mind, we designed and deployed the EvaluLLM to enable users to create evaluation tasks. The testing sessions were conducted via semi-structured interviews online through video conferencing application, Webex. Initially, the researcher provided participants with a link to the research prototype, which they were asked to access and then share their screen content. After providing an overview of the study's focus and tasks, the participants proceeded to create their evaluation tasks and share their thought processes via thinking-aloud methods. All participants were given the same task, which was to define LLM-as-a-judge to help find the best model for generating a good headline from the CNN/Daily Mail dataset.

%\textcolor{red}{Left align the table}

\subsection{Participants}
We recruited 8 industry professionals (Appendix Table \ref{tab:my_label}) with deep domain knowledge in model evaluation at a large technology company (2 females and 6 males) via social media recruiting, with participation and recommendations from various individuals. These industry professionals primarily consist of data scientists, software engineers, and AI engineers. Eligible participants were those who had hands-on experience evaluating large language model performance in their projects in the past year. The interviews were conducted remotely, and participants volunteered and consented to the recording of the session, as well as to the use of the interview results for research purposes.

% \subsection{Data Analysis}
% We began our analysis of the semi-structured interviews by carefully transcribing the recordings. To ensure accuracy, we examined the auto-transcribed transcripts while re-watching the interview recordings, capturing every nuance in participant responses. Following the transcription, we embarked on a thematic analysis. This process involved two researchers working collaboratively yet independently to examine the transcripts, aiming to identify emergent themes and underlying patterns. Our coding strategy was twofold: deductive, grounded in predefined themes derived from our research questions and inductive, allowing new themes to arise directly from the data. The thematic categorization involved a detailed process of extracting and grouping participant quotes based on thematic similarities. These groupings were further organized into three overarching categories that aligned with the core aspects of our research questions: evaluation use cases, criteria, and workflows.

% This approach facilitated a thorough analysis, wherein themes were systematically identified, coded, and examined to build a nuanced understanding of user needs and challenges related to our study's core areas.. This detailed and systematic process enhances the trustworthiness and credibility of our findings, yielding insightful design recommendations for LLM-as-a-Judge design

\subsection{Data Analysis}
Two authors independently reviewed the transcripts from recorded video sessions to pinpoint users' needs, system shortcomings, and challenges in the evaluation workflow. This independent review helped minimize bias and allowed for a comprehensive data exploration. Each author used a codebook of example quotes to support the identified themes. The authors then met to merge similar themes and address any initially missed, resulting in three main categories: use case challenges, evaluation criteria, and evaluation workflow, detailed in Appendix Table \ref{tab:results}. This classification captures the complexities of the evaluation process, encompassing users' needs, system limitations, and evaluative challenges.

\section{Results}
Our data analysis identified nine themes, categorized into use case challenges, evaluation criteria, and evaluation workflow (for a full list with example quotes see Table \ref{tab:results} in the Appendix). 

\subsection{Use Case Challenges}
The system requires users to input a prompt for their specific task, after which it generates the output and proceeds with the evaluation. This approach involves sending the identical prompt to various models for output evaluation. However, this methodology poses limitations for experienced users who tailor prompts for specific models, such as LLaMA.  Our participants described instances of \textbf{absence of specifications} where clients lack clarity on the task's data requirements. 

%``And I think there are 2 ways of doing it once more. It is when the client gives us the, the ground truth, or the kind of answer they expect from the model. So we can compare using, uh, metrics such as or BLEU, And this is like this other scenario, which unfortunately is more common, which is client doesn't even know what they want.'' (P5)
%When starting a project, it is common that a developer has to develop and evaluate the solution without having reference data that reflects their client’s preferences or requirements. For example, when inquiring about their current experience with model evaluation, P5 categorizes evaluation into two broad categories,
%\begin{quote}
%\textit{
%\end{quote}

%When asking about the percentage between these two scenarios, \textit{"it was like eighty-twenty, eighty percent of the time they don't have it."} However, this situation might change as the project progresses, and more teams co-create reference data with the client, making the reference data available in an accumulated fashion. 

%Since, ultimately, more and higher-quality reference data will become available, enabling model evaluation with a few-shot example or more reference-based methods, such as BLEU or ROUGE to compare.  This requires us to design a more flexible LLM evaluation system to overcome the cold start challenge with no reference data at the beginning while accommodating the changes and taking advantage of those reference data during the process. 

%\subsubsection{Support Rapid Performance Comparison with different configurations}
Additionally, there are numerous open-source and closed-source LLM models available, and users would like to test various setups, e.g., model selections, model configurations, and prompts. They would like the system to \textbf{support comparison with different setups.} Given time constraints and limited investment resources, it is often impractical to test all models with their use case data.
Teams usually begin with top-performing models, either from public benchmarks close to their use case or chosen based on their well-known reputation. Model selection is transient and highly constrained by project requirements. Instead of evaluating multiple models' performance with different prompts, they typically start with 1-2 models and improve performance through prompt engineering. This involves running the model with various prompts and parameter settings, where they often iterate over the setup to match specific baseline performance. It requires rapid performance comparison and support for evaluation data, accommodating multiple models and considering combinations with different setups.
%``We think we would take a bunch of candidates say we had five different models and for each model we had 20 different configurations or something like that. Now that's 100 different combinations. Um, we'd like the limited judge to be to run on like all hundred Give us an overview…which are the three that are actually worth looking at?'' (P2)
%\end{quote}

%For example, P7 \textit{"I mean, more so from like, a selection perspective… it's not even worth our time to look at some of these other models that are not at the top."} and \textit{"So we just try to truncate the models upfront to what we think is the best."}

%, for example, P7 states, \textit{"GPT 4 as a baseline and we're just trying to see how close are we getting with these other models in order to replicate the performance."}

\textbf{Shifting evaluation priority }often occurs as the project progresses. At the beginning of the project, where the main purpose is often the proof of concept for a specific proposed solution, the evaluation focus is mainly around feasibility testing. This involves assessing whether the proposed system or solution can produce accurate answers. However, as the project progresses into production, the evaluation purpose might shift from rapid model performance comparison to continual improvement with user feedback, performance monitoring, and reporting potential issues to draw developers' attentions. As evaluation priorities might differ for various use cases in different project phases, when designing an LLM-as-a-Judge solution, shared needs among these different phases and unique requirements in each phase need to be clearly articulated. This could help better define and design the experience and interaction to effectively support the diverse requirements for each phase.

%\begin{quote}
 %\textit{"I know that's like a terrible metric [confusion matrix] to be used as the first one, but we have actually done this with a client because they asked us to do so. They're looking for just accuracy."} - P5
 %\end{quote}
% \textit{"we would be willing to sacrifice a little bit of latency if the output is superior, and additionally for us, you know, the variable cost associated with processing is almost irrelevant when it comes to the things that we do for…this is something that can change in the production environment, but when we're trying to get the business, all we care about is performance."}

\subsection{Evaluation Criteria}
We identified several themes related to how users developed, changed, and trusted the evaluation criteria they were working with. While participants appreciated the flexibility of using the freeform approach in EvaluLLM, many expressed that they \textbf{desire structured and customizable templates} for specific use cases that can be tweaked for their purposes. They believe such templates would help them start with an evaluation baseline. 

Moreover, participants highlighted the necessity of distinct evaluation criteria for various tasks. For example, they noted that a RAG task might require one set of criteria, while a creative task might demand another. Participants often crafted criteria complete with descriptions and scoring. One typical approach involved naming each criterion, defining it, and then assigning a score.

Evaluation criteria serve as a medium to communicate user preferences to the model. An effective criterion not only needs to reflect the user’s preferences but also must function well to enable the model to understand and follow instructions. When reflecting on evaluation criteria, participants expressed the \textbf{need for multiple rounds of iterations} when refining their criteria. \textit{"It can be really hard to figure out how to express the evaluation criteria in a way that makes sense to the model. But it can also just be hard in your own mind to figure out what it means for a title to be good." P2 }

%\end{quote}
%\begin{quote}
%\textit{"It can be really hard to figure out how to express the %evaluation criteria in a way that makes sense to the model. But it can also just be hard in your own mind to figure out what it means for a title to be good."} - P2 
%\end{quote}

The importance of giving supporting multiple rounds to refine and expand criteria emerged when looking at the types of dimensions participants created. We found that users tend to prioritize more objective metrics such as accuracy before they start to consider the styling of the outcome. At the beginning of the project, the primary concern for a client is getting the correct answer from the model. That is not to say, that our participants did not care about more subjective criteria, but that happens later in the process. 

%It is possible to establish a universal definition of accuracy and requirements for the output length. However, the term "punchy" poses a challenge, given its subjective nature and the potential for diverse personal opinions and preferences.

%Throughout our study, \textit{"I think that's the biggest concern, and so coming up with wrong answers, uh"}(P5) and \textit{" maybe do it [consider secondary list of dimensions] depends on our client's maturity. It's pretty much go or no-go for [correct and incorrect answer]"} (P5)

Although users might have a rough idea of what they want, it is challenging to describe everything at the beginning, especially when they don’t have access to the evaluation data. One participant struggled during the criteria definition process as he was required to define the criteria before he could see the output data. Providing the output might help users articulate what they want or don't want, assisting them in iterating the criteria description or adding examples to better align with their preferences. 

%``So I've created a template here, maybe could show an example of here's the problem with this, this with a sample article, plug it in so I can see what it looks like.'' (P2)
%\end{quote}

%\textit{"What I really want is the ability to say, hey, this part of the output is great. Let me add that as an example or include that as part of the evaluation criteria in one of these"} (P3)
Users express a desire for more than just a high-level result summary; they are keen on obtaining a detailed breakdown of each dimension and a need for the system to \textbf{display performance for each criteria individually}. EvaluLLM currently only presents a win rate as a high-level performance summary metric to showcase the winning model on the leaderboard. Participants expressed the desire to view performance across each dimension rather than a high level win-rate. 

% \textit{"Win rate is not necessarily the best metric because there are multiple categories to define what it means to win." (P7)}

%\textit{"So I'm covering a lot of ground there, and I know that's hard for the model to deal with because now the model has to have a whole lot of different criteria, and it's all drawn up by the ones, but that's kind of what a good title headline is about."}

 %P7 further emphasized,
 %\begin{quote}
  %\textit{"
  %\end{quote}

    % Besides the breakdown of performance for each dimension, participants stress the importance of relative difference, as P1 elaborated on this, stating, "\textit{Relative difference is important to me - between the different evaluations. I typically rank them. Hey, I've got 1, 2, 3 - are we just splitting hairs, or is it NO 1 is head and shoulders above 2 and 3? There are also other considerations - I always want to position \textcolor{red}{granite} because of all the things I know and love about granite, but if it sucks, I can’t, right? If it's 2 behind llama, I'm good to go - but if it's 50\% below, then I am not even gonna bother."} \textcolor{red}{should we mask model name when we include the quote with this info?}

%\textcolor{red}{should this be another theme?}

\subsection{Evaluation Workflow}
While presenting the tool to users, we probed them on their current evaluation workflows and how they would imagine incorporating EvaluLLM. Users expressed the challenges they faced when doing manual evaluations and how they would use automated methods and the EvaluLLM experience to address those challenges. Although there are only 10 examples in our testing dataset, generating the evaluation results after user created the evaluation is time consuming because of calls to the model. Model calls are expensive and time consuming and one potential way to address this is to \textbf{run the evaluation on a subset of the data first}. 
% As expressed by P1, \textit{"I’d want to iterate on my judge enough for it to get a decent annotator agreement and then let it go wild."} 

%``We don't have a problem here because the data set is small. But, like, if there's like, a 1000. Then it would it make sense to go through the entire batch and we find out your volume criteria needs to be tweaked'' (P7)
%\end{quote}

%Instead running the auto-evaluation with the complete dataset, users prefer to a more light weight experiment with a small subset of data to refine the criteria before they kickoff the full automation, 
%\begin{quote}
%\textit{"It might make sense to have a button, say try it out on a couple examples and then you get you get to see what it's doing immediately before having to run on the full dataset and see what it does on everything"} - P2
%\end{quote}
%\begin{quote}
%\textit{"} - P1 
%\end{quote}

%\subsubsection{Instant feedback provides more autonomy}
To evaluate the agreement of the LLM-as-a-Judge preferences with humans, participants were asked to conduct blind reviews of the model's output. These reviews would be utilized to calculate the agreement between the LLM-as-a-judge and the participants. While it is beneficial to observe the agreement rate in the summary page, users also desire more control over the workflow and seek instant feedback during the manual review process. They would like to see how much the LLM-as-a-judge agrees with them once they provide feedback and wish for the system to proactively provide criteria modification suggestions. One way of providing \textbf{instant feedback on human-AI agreement} is to allow users to either initially upload human evaluations for comparison with the automatic evaluations. Another way is to conduct a blind review before the evaluations are presented, ensuring that users receive instant feedback on human-AI agreement as soon as the evaluations are ready.
%\textit{"tell me when to quit"}(P1)

During testing, we observed that some participants might provide overly detailed instructions for both the task prompt and the evaluation criteria. The design intention was to simplify the user input requirements, seeking only the evaluation criteria rather than a complete evaluation prompt with detailed evaluation process. However, some participants included the step-by-step evaluation process in the criteria definition input. Additionally, some participants inquired about adjusting their evaluations per judge.
 % \textit{"I know when I'm using llama, I typically put a bunch of these weird angle bracket tags and bracket insert and stuff like that. are you guys doing that for me or should I be doing it?" (P2).}
 
As our participants are domain experts in model evaluation, they are well aware of potential biases in the model. They actively seek transparency regarding the bias mitigation strategy to effectively calibrate their trust in LLM-as-a-Judge results. Additionally, participants were cognizant of self-enhancement bias \cite{zheng2023judging} and expressed concerns about the LLM-as-a-judge being one of the models to be evaluated. \textbf{Ensuring transparency for trustworthy evaluation} was deemed crucial by users, such as transparency concerning the prompts sent to the judge and whether bias mitigation has been implemented. One user remarked, \textit{"It seems like Granite always displays first, and Flan-UL-2 always comes second. Does the system randomly switch positions?" P5}

\subsection{Limitations}
Our study is based on a small sample of only 8 domain experts, potentially impacting the generalizability of our findings. In addition, our methodology primarily concentrated on observing users utilizing our specific evaluation tool with one pre-defined dataset. This approach may restrict the broader applicability of our results. Note that EvaluLLM at the time of this study was a functioning proof-of-concept but not yet a scalable systems that can be deployed to a large user population.  However, we believe our findings still offer relevant insights into the challenges and needs users encounter when using LLM-as-a-Judge tools, as evidenced by our focused line of questioning aimed at understanding how more automated evaluations integrate into users' workflows.

% Our study has several limitations. First, our recruitment process involved employees who mainly interact with clients rather than end consumers, potentially impacting the generalizability of our findings. Secondly, our methodology primarily focused on observing users using our specific evaluation tool, which may limit the broader applicability of our results. However, we believe our findings still offer relevant insights into the challenges and needs users encounter when evaluating LLM-as-a-Judge tools, as evidenced by our focused line of questioning aimed at understanding how automatic evaluation integrates into users' workflows.

\section{Discussion and Design Recommendations}
Our findings highlight user needs across different use cases when using LLM-as-a-judge. Users require guidance to evaluate model outputs effectively. We discuss the implications of our findings and propose design recommendations for LLM-as-a-judge tools and user experiences. 

% \textcolor{red}
% {reflect on the outlined limitations of our design recommendations and outline future work and design directions to further validate our design recommendations and improve writing.}

%\subsection{Design Recommendations}

\begin{figure*}[t]
\centering
\includegraphics[width=0.9\textwidth]{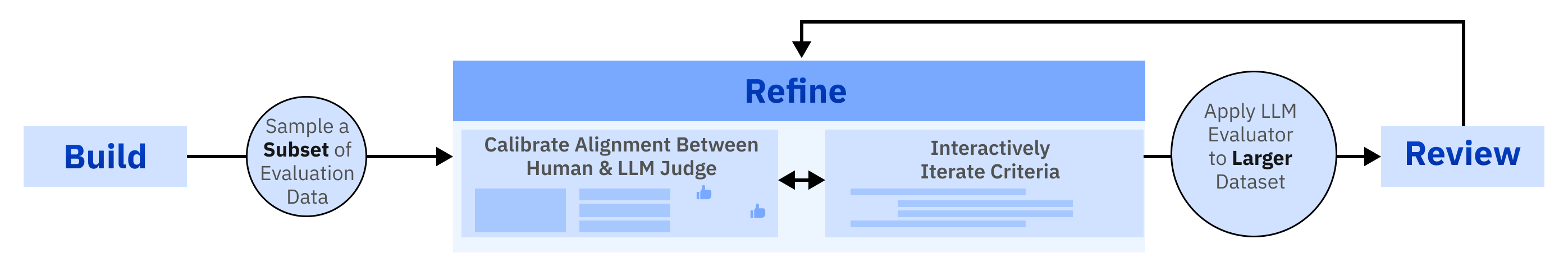}
\caption{Recommended evaluation workflow: interactive refinement of criteria with a subset of data prior to applying evaluation to entire dataset can potentially improve preference alignment and trust calibration.}
\label{fig2}
\end{figure*}

\subsection{Efficient Criteria Iteration}
LLMs can generate high-quality outputs aligned with human preferences, but processing the entire dataset is costly and time-consuming, especially with methods like pairwise comparisons, which increase compute costs significantly. To optimize efficiency, it's advisable to start a project by allowing users to refine their evaluation criteria using a representative data sample before scaling up to the full dataset (see Figure \ref{fig2}). Effective sampling enhances learning for LLM-as-a-Judge by selecting diverse and representative outputs. Techniques like clustering \cite{chang2021training} or graph-based search \cite{su2022selective} can aid in output selection for human evaluation. Addressing misalignments and manually reviewing low-confidence outputs \cite{desmond2021semi} are crucial, as is displaying a subset of evaluations to lessen users' cognitive load and facilitate iterative refinement of evaluation criteria.

\subsection{Structured and Customizable Templates}
For creative generation tasks, it's crucial to employ diverse, custom criteria. To streamline this process, we propose providing standard criteria that are universally applicable across various use cases, supplemented by customizable templates. As illustrated in our design explorations (see Appendix Figure \ref{fig3}), users can select from predefined criteria dimensions (Figure \ref{fig3}A) or utilize recommended templates for common scenarios (Figure \ref{fig3}B). These templates are designed to be flexible, allowing easy adaptation to specific user needs.

Further enhancing customization, the proposed templates support hierarchical organization (see Appendix Figure \ref{fig4}), enabling the addition of new criteria dimensions (Figure \ref{fig4}G), nesting of sub-criteria (Figure \ref{fig4}F), and removal of unwanted elements (Figure \ref{fig4}H). Users can also adjust scoring scales (Figure \ref{fig4}E). This hierarchical structure, supported by findings from related works \cite{zheng2023judging} \cite{kim2023} \cite{stureborg2023interface}, allows users to start with broad criteria and refine them to capture specific task nuances. To foster ongoing improvement and reuse, the system should enable users to save and share these templates (Figure \ref{fig4}B). Considering the benefits of balanced evaluations, users should be able to adjust the weight of different criteria dimensions, aligning more closely with human preferences. The inclusion of reference examples within the templates (Figure \ref{fig4}D) can further refine the criteria based on actual output data, enhancing the preference agreement process. This approach not only makes the criteria definition process more efficient but also ensures consistency and rigor in evaluating creative tasks, leading to more accurate and effective assessments.

Providing structured and customizable templates will not only expedite the process of criteria definition but also foster consistency and rigor in the evaluation of creative generation tasks, which will contribute to more accurate and effective evaluations.

\subsection{Interactive Criteria Iteration}
Our findings revealed crafting effective criteria typically requires multiple iterations. Criteria components such as name, definition, scale, and examples often need definition and refinement as users evaluate outputs. Users include examples of both poor and excellent outputs to help LLM-as-Judges distinguish quality through few-shot learning techniques. Related work \cite{kim2023} indicates that users often develop new criteria during evaluations. To facilitate this process, a real-time feedback system that allows users to immediately see the impact of criteria modifications would be useful. Additionally, a user-friendly interface that enables easy modification and experimentation with criteria could significantly improve the efficiency and customization of the evaluation process.

% \textcolor{red}{Need to add a Figure to show interactive criteria definition}

\subsection{Ensure Consistency} As human preferences may not be consistent within the same set, aligning with frequently changing preferences becomes a challenge. A self-consistency check mechanism can expedite this alignment. When refining criteria, any discrepancies between human and LLM-as-a-Judge evaluations should prompt a review of similar sample data post-calibration. Incorporating an automated consistency checker that flags potential criteria conflicts or inconsistencies could streamline the evaluation process by offering actionable solutions to address these inconsistencies. Leveraging the diversity of logical paths in complex reasoning tasks, as suggested by recent studies \cite{Stanovich_West_2000}, the self-consistency CoT method \cite{wang2023selfconsistency} can generate multiple reasoning paths, selecting the most consistent answers by averaging over these paths, thus improving evaluation outcomes.

\subsection{Support Different Setups}
Our findings emphasize the need for an LLM to function flexibly as a judging system throughout different project phases. It should support a variety of evaluation data configurations, including diverse model selections, prompts, and settings. While some evaluations may only compare outputs from a specific prompt and model setting, optimal performance often requires tailored prompts and settings for each model, involving substantial prompt engineering and comparison of different configurations. Thus, the system must not only evaluate common settings across various models but also assess various prompts and settings for select models, highlighting the importance of designing an adaptable LLM judging system.

\subsection{Adaptable Reference-Based Evaluation}
Our user study findings showed that users often start projects without clear objectives, resulting in evaluations lacking reference data. Users interacting with the LLM-as-a-Judge system gradually accumulate reference data, either directly or from external sources, so it could be beneficial to design systems that incorporate human input to refine preference correspondence using expert-labeled data \cite{liu2023calibrating} or other collected references. This flexible approach enhances the system's effectiveness and trustworthiness, ensuring it evolves in line with user preferences.

\subsection{Enhance System Transparency}
Our findings indicate that users value transparency to comprehend the LLM's role as a judge. This encompasses access to essential details like the specific prompt used (illustrated in Figure \ref{fig5}A) and the implementation of bias mitigation strategies. To design an effective LLM-as-a-Judge system, it is critical to make such information readily available. This can be facilitated by allowing users to view the prompt, enabling the system to explain evaluation results, and integrating visualization tools that demonstrate how user inputs affect the evaluation process.
%Based on our learning, users seek transparency to better understand the LLM as an judge. This includes details such as the actual prompt (see design example in Figure \ref{fig5}A) sent to the system and whether bias mitigation measures have been implemented. In designing the LLM-as-a-Judge, we should ensure this information is easily accessible to users. This can be achieved by enabling users to view the prompt, having the LLM-as-a-Judge explain their evaluation results, incorporating visualization tools to help users understand how their inputs impact the evaluation process.

% , and even allowing for self-critique of misalignments with reasoning via the Chain of Thought (CoT) technique \cite{wei2022chain} for further improvement.

\subsection{Proactively Mitigate Potential Bias}
%\textcolor{red}{Given the importance and persistent challenges of mitigating bias, it's crucial to consistently implement strategies for fairer evaluations in LLM-as-judge systems, including, but not limited to, methods such as .
Considering the persistent challenge of bias, systems should implement bias mitigation strategies that include swapping answer order to reduce position bias \cite{zheng2023judging} and treating inconsistent results as ties, or by randomly assigning positions in large datasets \cite{li2023alpacaeval} \cite{zheng2023judging}. For verbosity bias, the "repetitive list" attack technique \cite{zheng2023judging} challenges LLMs to favor clarity over length in responses. Furthermore, enhancing LLMs' abilities in mathematical and reasoning tasks can be achieved through Chain-of-Thought approaches \cite{wei2022chain}, coupled with reference-guided evaluation where the LLM generates and then evaluates its own initial responses.

\subsection{Explore Further Automation}
Our study found that task prompts often contain criteria, suggesting the possibility of extracting them automatically for tailored guidelines. Related work also shows that users prefer automated prompt refinement over manual revisions \cite{kim2023}. Various suggestions(see Appendix Figure \ref{fig5}), such as rephrasing (Figure \ref{fig5}A), adding reference examples (Figure \ref{fig5}B), incorporating more scales (Figure \ref{fig5}C), and introducing additional dimensions (Figure \ref{fig5}D), could be proactively provided by the system for humans to review to further accelerate evaluation correspondence. While these areas show promise for further improving the efficiency of preference correspondence, considering the limitations of automation systems, it is essential to place humans in the loop to calibrate accuracy and trustworthiness.

\section{Conclusion}
We studied EvaluLLM, an AI-assisted tool utilizing LLMs alongside humans as judges for LLM-generated content. Our findings highlight the potential of LLMs as customizable judges and underscore the importance of interactive, transparent, and user-centered evaluation processes. Based on our findings, we offer design suggestions for practitioners that can help them build more effective , nuanced, adaptable, and user-friendly evaluation tools that meet diverse needs as compared to automated benchmarks. Inspired by our user research, we are currently in the process of rolling out an evolved AI-assisted evaluation tool to a larger user population to observe "usage in the wild."

%Our recommendations include proactive measures to reduce bias and enhance fair evaluation. Ultimately, we aim to enhance NLG quality assessment by combining machine efficiency with human-centered approaches.

% Our study describes EvaluLLM, a tool to use large language models (LLMs) as judges for natural language generation (NLG) outputs. Our findings highlight the potential of LLMs as customizable judges and underscores the importance of interactive, transparent, and criteria-driven evaluation processes. We provide design recommendations aim to develop more nuanced, adaptable, and user-friendly tools to meet the diverse needs of practitioners. Ultimately, our work contributes to improving NLG quality assessment, advocating for integrating machine efficiency with human-centric approaches.

%\bibliographystyle{ACM-Reference-Format}
\bibliography{custom}

\appendix
\section{Participant Information}
Table \ref{tab:my_label} shows the details of participants involved in the user study, predominantly comprising of industry experts such as data scientists, software engineers, and AI engineers. These professionals have practical experience in evaluating the performance of large language models in their projects over the last year.

\begin{table*}[b]
    \centering
    \begin{tabular}{lll}
     \textit{ ID}  & \textit{Gender }& \textit{Job Role} \\
      \hline
       P1 & Male & Lead Software Engineer/Data Scientist \\
      \hline
       P2 & Male & Principle Data Scientist  \\
       \hline
       P3 & Male & Lead Software Engineer/Data Scientist \\
       \hline
       P4 & Male & Data Scientist  \\
      \hline
       P5 & Male & AI Engineer/Data Scientist \\
       \hline
       P6 & Female & Data Scientist \\
       \hline
       P7 & Male & Senior Technical Manager/Data Scientist  \\
       \hline
       P8 & Female & Data Scientist \\
                    
    \end{tabular}
    \caption{Demographic information from participants in our user study.}
    \label{tab:my_label}
\end{table*}

% \section{User Study:Interview Questions}
% \begin{itemize}
%     \item \textit{Participant Background}: "What is your main job role and if you have any related experience around large model evaluation for projects"  
%     \item \textit{Evaluation Use Case}: "What kind use case you usually evaluate models for? "  
%         \item \textit{Think Aloud Exercise}: "I am going to gather your help to try out our latest Evaluation tool? here is the link and here is the task "  
% \end{itemize}

\section{Summary of Evaluation Themes and Examples}
Table \ref{tab:results} provides further details on evaluation themes generated from the user study, along with corresponding examples from participants' quotes.

\begin{table*}[ht]
\centering
\small
\caption{Table of evaluation themes and corresponding examples. Themes are grouped into three categories: use case challenges, evaluation criteria, and evaluation workflow. Quotes are provided to delineate themes. }
\label{tab:results}
\begin{tabular}{p{2.8cm}p{3.5cm}p{9.2cm}}
\toprule
\textbf{Group} & \textbf{Theme} & \textbf{Example} \\
\midrule
Use Case Challenges &  Absence of Specifications &
\textit{``So we can compare using, metrics such as or BLEU, And this is like this other scenario, which unfortunately is more common, which is client doesn't even know what they want.''} - P5\\\\
&&
\textit{``It was like eighty-twenty, eighty percent of the time they don't have it.''} - P5\\\\
&Support Comparison with Different Setup&
\textit{``Say we had five different models and for each model we had 20 different configurations or something like that. Now that's 100 different combinations. Um, we'd like the limited judge to be to run on like all hundred. Give us an overview. Which are the three that are actually worth looking at?''} - P2\\\\
&&
\textit{``GPT 4 as a baseline and we're just trying to see how close are we getting with these other models in order to replicate the performance.''} - P7\\\\
&Shifting Evaluation Priority& \textit{``I know that's like a terrible metric [confusion matrix] to be used as the first one, but we have actually done this with a client because they asked us to do so. They're looking for just accuracy.''} - P5  \\\\
&&
\textit{``GPT 4 as a baseline and we're just trying to see how close are we getting with these other models in order to replicate the performance.''} - P7\\\\
%“we would be willing to sacrifice a little bit of latency if the output is superior, and additionally for us, you know, the variable cost associated with processing is almost irrelevant when it comes to the things that we do for…this is something that can change in the production environment, but when we're trying to get the business, all we care about is performance.”- P7 \\
%`So, and I think also the nice thing about this is that it's very flexible. Mm-hmm. So if I want to say, in this kind of headline generation, right, maybe I want to give extra credit if it's a clever headline in some sense, right? There's a ponder, maybe there's a rhyme, right? That kind of creativity. (P3) 
%"More examples might be nice (P2)
Evaluation Criteria & Desire Structured and Customizable Templates & 
\textit{``A freeform text box is too simple. I would love there to be templates that I can utilize. And at the very least, be able to just edit so that I can get into my use case.''} - P7 \\\\
&&
\textit{``More examples might be nice.''} - P2\\\\
% &&
% \textit{``So I do like having this freeform evaluation criterion, but I think maybe there being an option for it to be, here's a suggested criterion that's good for some use cases. And if you want it to be different, you can.''} - P3\\\\
% &&
% \textit{``Maybe we can just, like, click and select those criteria to be added to the prompt of all the way the evaluation prompt. ''} - P5\\\\
& Need for Multiple Rounds of Iterations & \textit{``It can be really hard to figure out how to express the evaluation criteria in a way that makes sense to the model. But it can also just be hard in your own mind to figure out what it means for a title to be good.''} - P2 \\\\
&&
\textit{``If I think, without having a clearer sense of what the evaluation is, sort of what a baseline evaluation is, it might be nice to have a couple of features of an evaluation that we could just select in like a checkbox. ''} - P3\\\\
%“But I don't know how to calibrate that evaluation to any intuition that I might have about the output.” P3
%What I really want is the ability to say, hey, this part of the output is great. Let me add that as an example or include that as part of the evaluation criteria in one of these' (P3)\\
%“So I'm covering a lot of ground there, and I know that's hard for the model to deal with because now the model has to have a whole lot of different criteria, and it's all drawn up by the ones, but that's kind of what a good title headline is about.” P2
&Display Performance for each Criteria Individually&\textit{ ''There might be times where you have to trade off on certain kinds of things and Win rate is not necessarily the best metric because there are multiple categories to define what it means to win.''}  - P7 \\\\
&&
\textit{``So I'm covering a lot of ground there, and I know that's hard for the model to deal with because now the model has to have a whole lot of different criteria, and it's all drawn up by the ones, but that's kind of what a good title headline is about.''} - P7\\\\

Evaluation Workflow&Run Evaluation on Subset of Data First & \textit{``We don't have a problem here because the data set is small. But, like, if there's like, a 1000. Then it would it make sense to go through the entire batch and we find out your volume criteria needs to be tweaked.''} - P2 \\\\
&&
\textit{``I’d want to iterate on my judge enough for it to get a decent annotator agreement and then let it go wild.''} - P2\\\\
&Instant Feedback on Human-AI agreement& \textit{``Tell me when to quit.'' }-P1\\\\
&Ensuring Transparency for Trustworthy Evaluation&
\textit{ ``So I definitely want, as we discussed earlier, a lot of transparency and exactly what is being sent to the models to generate the responses and then what is then being sent to the LLM as a judge.''} - P2\\\\
&&
\textit{``Maybe a small note on, like, you know what the prompt is, like, what the data set is and what the tool is doing.''} - P8\\
 %“I don't know how accurate it will be evaluating itself and how biased it will be because if it's generating the headline (P5)”\\

\bottomrule
\end{tabular}

\end{table*}

\section{Recommended Designs}
\label{sec:appendix}
\begin{figure*}[t]
\centering
\includegraphics[width=0.99\textwidth]{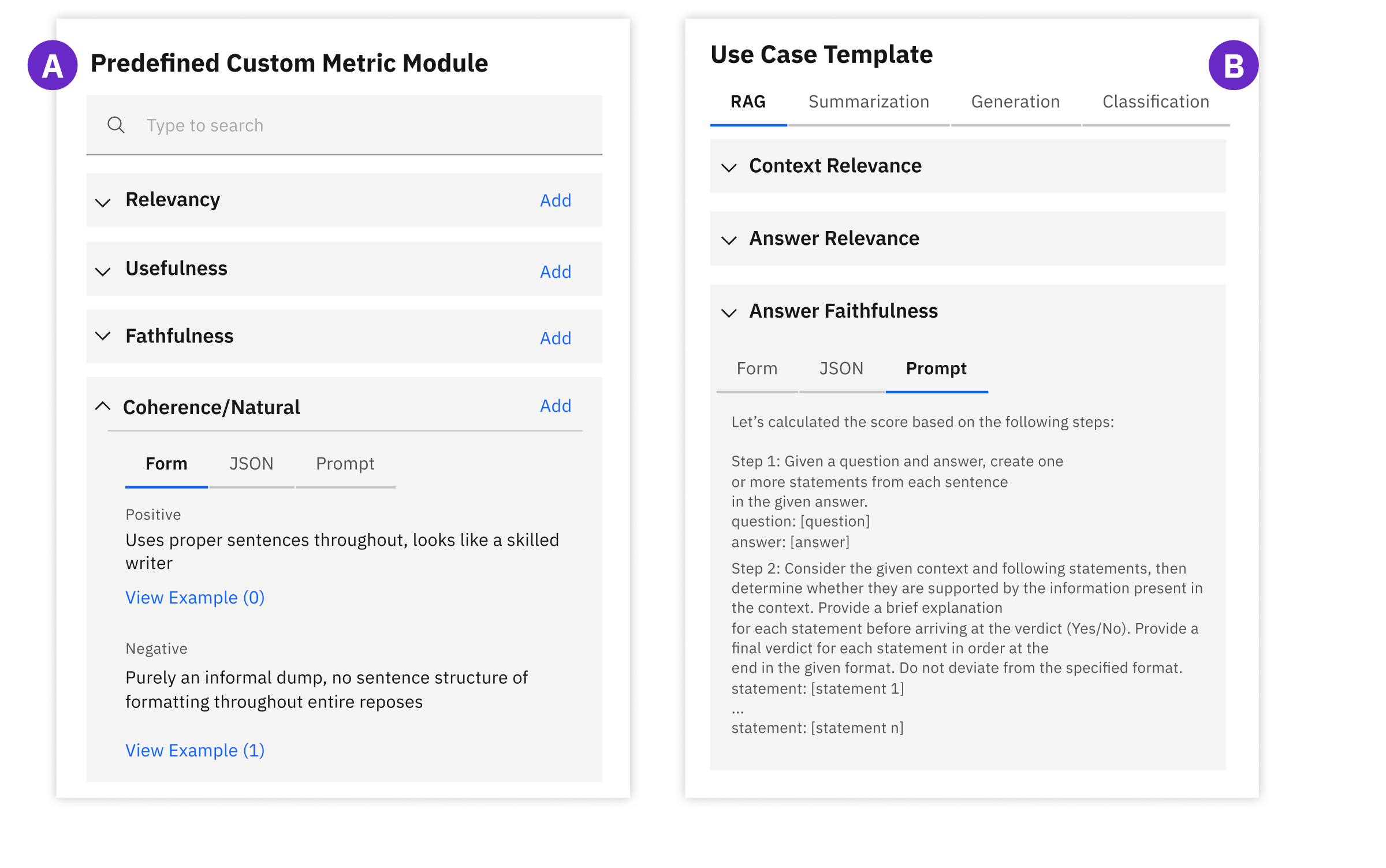}
\caption{Recommended design to (A) enable users to choose from a list of predefined custom metric modules and (B) enable users to create a set of evaluation criteria based on common use cases. }
\label{fig3}
\end{figure*}

\begin{figure*}[t]
\centering
\includegraphics[width=0.99\textwidth]{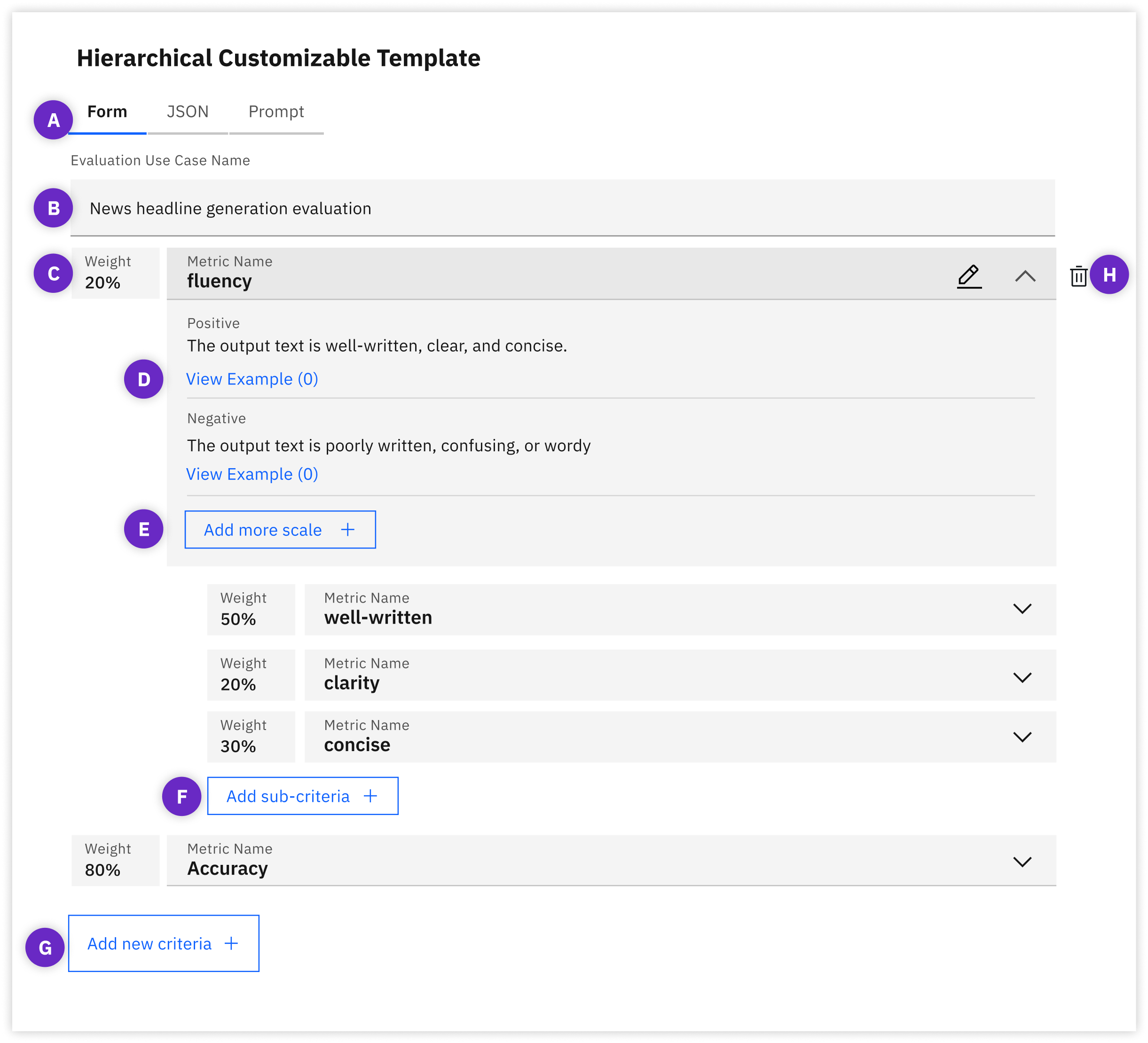}
\caption{Recommended design to provide structured and customizable templates that support hierarchical, multi-dimensional evaluations. }
\label{fig4}
\end{figure*}

\begin{figure*}[t]
\centering
\includegraphics[width=0.99\textwidth]{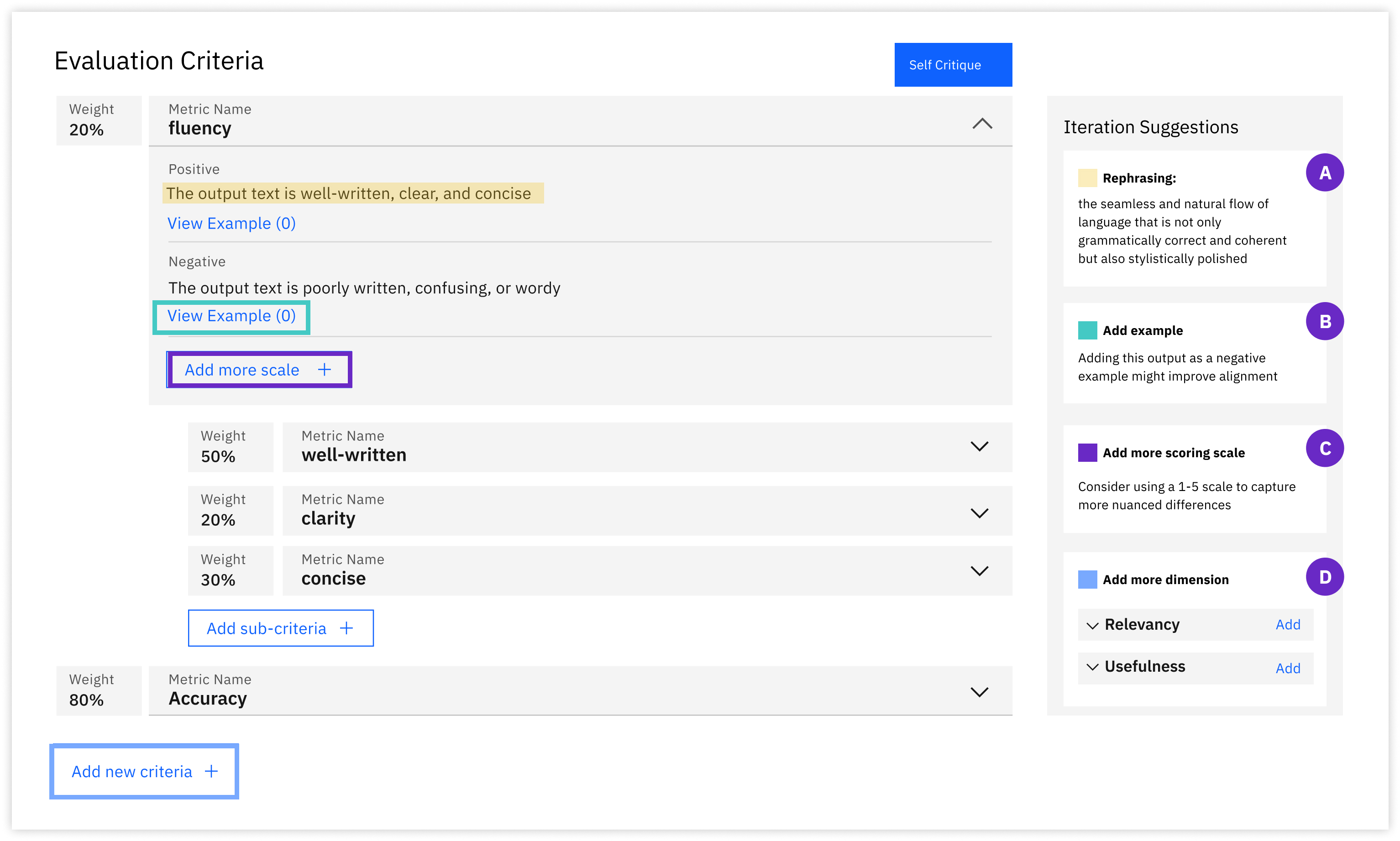}
\caption{Recommended design demonstrating the ability of users to leverage LLM-as-a-Judge for Criteria Iteration.}
\label{fig5}
\end{figure*}

Figure (\ref{fig3})(\ref{fig4})(\ref{fig5}) show design examples to help illustrate corresponding design recommendations.

\section{EvaluLLM Evaluation Workflow}
\begin{figure*}[t]
\centering
\includegraphics[width=0.98\textwidth]{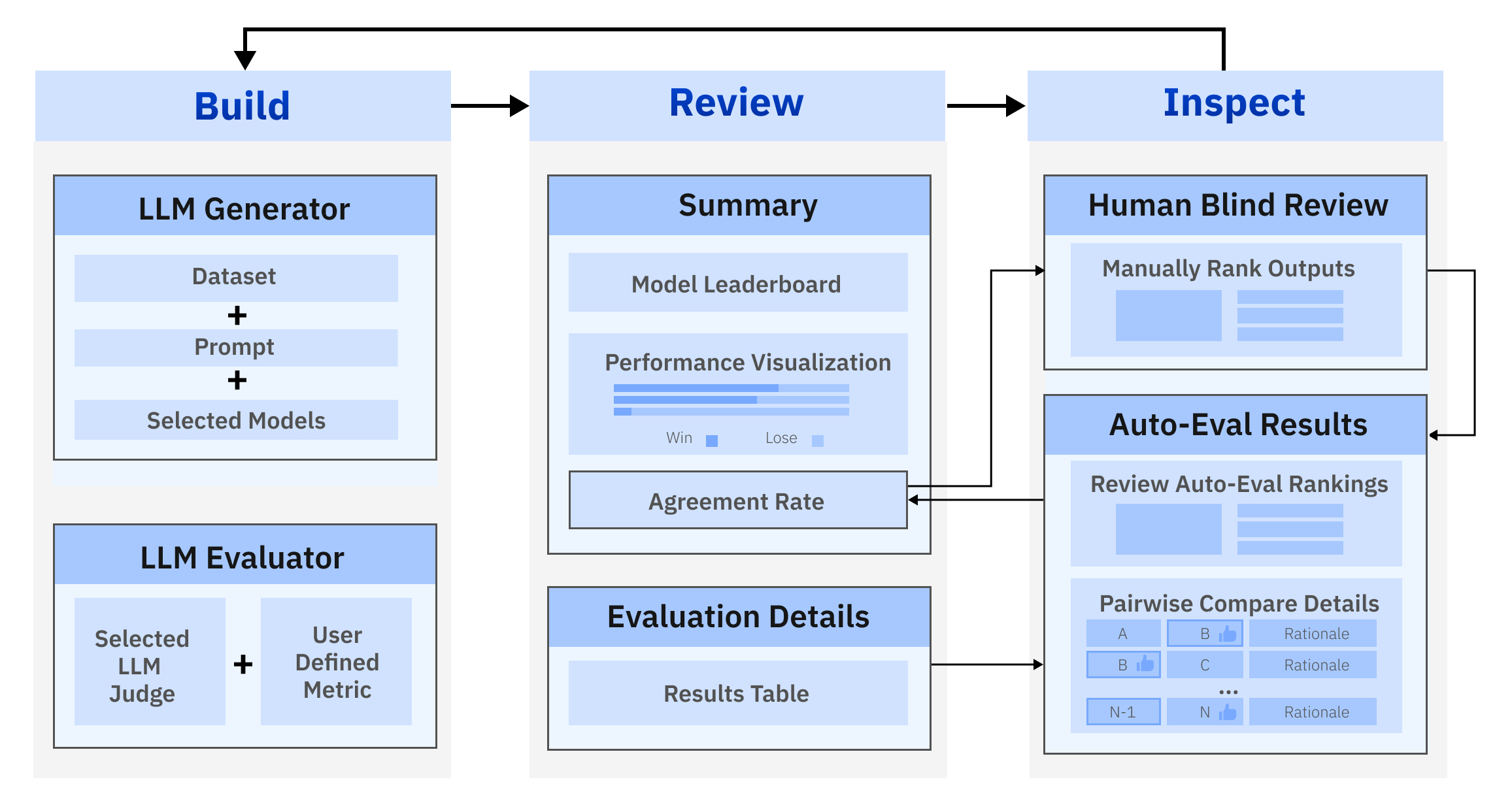}
\caption{EvaluLLM evaluation workflow overview which consists of a Build, Review, and Inspect process.}
\label{fig6}
\end{figure*}

Figure (\ref{fig6}) shows the high-level overview of the EvaluLLM workflow, which consists of a Build, Review, and Inspect process.

% \section{Implementation Details}
% \label{sec:appendix}
% \textcolor{red}{Need to verify if the following description and details are correct with team }

% We developed the front-end of EvaluLLM using TypeScript, ReactJS, and CSS. The back-end is built around a Flask server, integrating the Internal Foundation Model Platform Tool API to support all LLM components. For the LLM-as-a-Judge model, we chose meta-llam-2-70b-chat due to its superior human correlation rates compared to other models available for our research. Regarding the LLM-as-a-Judge configurations, we standardized the temperature setting to \textcolor{red}{specific value} across all components. Below are the full prompts used in EvaluLLM 

% <s>[INST] <<SYS>>
% You are a helpful and honest assistant.
% <</SYS>>

% This is an instruction:
% "{prompt}"

% There are two outputs to the instruction: 

% Output 1: "\{output_1\}" \\

% Output 2: "\{output_2\}"

% A quality output adheres to the following evaluation criteria:
% "{eval_{criteria}}"

% Which output is better quality based on the evaluation criteria?

% [/INST] Based on the evaluation criteria, the better quality output is output  

\end{document}